\documentclass[prd,twocolumn,showpacs,preprintnumbers,amsmath,amssymb,superscriptaddress,floatfix,nofootinbib]{revtex4}

\usepackage{graphicx}
\usepackage{amsmath}
\usepackage{amsfonts}
\usepackage{amssymb}
\usepackage{color}

\usepackage[colorlinks, citecolor=blue,anchorcolor=red,menucolor=red, linkcolor=red,filecolor=red,runcolor=red,urlcolor=blue,frenchlinks=red]{hyperref}

\begin{document}
\arraycolsep1.5pt

\title{Analysis of the $B^+\to J/\psi \phi K^+$ data at low $J/\psi \phi$ invariant masses and the $X(4140)$ and $X(4160)$ resonances}

\author{En Wang}
\affiliation{Department of Physics, Zhengzhou University, Zhengzhou, Henan 450001, China} 

\author{Ju-Jun Xie}
\affiliation{Institute of Modern Physics, Chinese Academy of
Sciences, Lanzhou 730000, China} 

\author{Li-Sheng Geng} 
\affiliation{School of Physics and Nuclear Energy Engineering and
International Research Center for Nuclei and Particles in the
Cosmos, Beihang University, Beijing 100191, China}

\author{Eulogio Oset}
\affiliation{Institute of Modern Physics, Chinese Academy of
Sciences, Lanzhou 730000, China}
\affiliation{Departamento de
F\'{\i}sica Te\'orica and IFIC, Centro Mixto Universidad de
Valencia-CSIC Institutos de Investigaci\'on de Paterna, Aptdo.
22085, 46071 Valencia, Spain}

\date{\today}

\begin{abstract}
We have studied the $J/\psi \phi$ mass distribution of the $B^+\to J/\psi\phi K^+$ reaction from threshold  to about 4250~MeV, and find that one needs the contribution of the $X(4140)$ with a narrow width, together with the $X(4160)$ which accounts for most of the strength of the distribution in that region. The existence of a clear cusp at the $D_s^* \bar{D}_s^*$ threshold indicates that the $X(4160)$ resonance is strongly tied  to the $D_s^* \bar{D}_s^*$ channel, which finds a natural interpretation in the molecular picture of this resonance.
\end{abstract}

\maketitle

The recent measurement of the $B^+ \to J/\psi \phi K^+$ reaction at LHCb~\cite{Aaij:2016nsc,Aaij:2016iza} and analysis of the data has brought some surprises as the claim of several new states that couple to $J/\psi \phi$. 
Another surprise is that the $X(4140)$ deduced from the analysis, with quantum numbers $J^{PC}=1^{++}$, however has a width $\Gamma \approx 83\pm 24^{+21}_{-14}$~MeV, substantially larger than the average $15.7\pm6.3$~MeV of the former experiments CDF~\cite{Aaltonen:2009tz,Aaltonen:2011at}, Belle~\cite{Brodzicka:2010zz}, LHCb~\cite{Aaij:2012pz}, CMS~\cite{Chatrchyan:2013dma}, D0~\cite{Abazov:2013xda,Abazov:2015sxa}, BABAR~\cite{Lees:2014lra}.

In this work, we show that the low invariant $J/\psi \phi$ mass region, poorly reproduced in the fit of Refs.~\cite{Aaij:2016nsc,Aaij:2016iza}, requires the contributions of a narrow $X(4140)$ and a wide $X(4160)$ resonance which couples to $J/\psi \phi$ but is mostly made by a $D_s^* \bar{D}_s^*$ molecule. As a consequence of this, the $J/\psi\phi$ mass spectrum develops a strong cusp around the $D_s^* \bar{D}_s^*$ threshold that the fit of Refs.~\cite{Aaij:2016nsc,Aaij:2016iza} cannot reproduce because this mechanism is not contemplated in the amplitude that is fitted to the data.

The $X(4140)$ has been the subject of much theoretical work and multiple suggestions have been given about its possible structure~\cite{Chen:2016qju,Guo:2017jvc}. Here we briefly discuss the works where claims of a $D_s^* \bar{D}_s^*$ molecule have been made. Effective Lagrangians accounting for pseudoscalar and vector exchange are used in Ref.~\cite{Liu:2009ei}, together with triangle diagrams to connect $D_s^* \bar{D}_s^*$ to $J/\psi\phi$. A state that could be associated with $X(4140)$ was obtained with $J^{PC}=0^{++}$ or $2^{++}$. In Ref.~\citep{Branz:2009yt}, a different strategy is followed involving  the 
Weinberg compositeness condition~\cite{Weinberg:1965zz,Baru:2003qq} and getting the couplings from the binding energy of $D_s^* \bar{D}_s^*$. Once again $0^{++}$ or $2^{++}$ are the preferred quantum numbers. In Ref.~\cite{Chen:2015fdn}, vector meson exchange, together with the Bethe Salpeter equations are used and once more a $0^{++}$ structure is favored. The possibility that the $X(4140)$ is a $D_s^* \bar{D}_s^*$
 molecule is also discussed in Ref.~\cite{Karliner:2016ith}, where $\eta$ exchange is used to connect heavy mesons which have no $u$, $d$ quarks, although the authors admit that other ingredients apart from $\eta$ exchange would be needed to bind that state. QCD sum rules have also contributed to this discussion and in Refs.~\cite{Albuquerque:2009ak,Zhang:2009st}, although with uncertainties of about 100~Mev in the mass, the possibility that the $X(4140)$ corresponds to a $D_s^* \bar{D}_s^*$ molecule with $0^{++}$ is pointed out. A study of this state using $\eta$, $\phi$, and $\sigma$ exchange is done in Ref.~\cite{Ding:2009vd}, concluding that the $X(4140)$ could correspond to a $D_s^* \bar{D}_s^*$ molecule with $0^{++}$ although $0^{-+}$ or $2^{++}$ were not excluded.
 
A through study of $X$ states emerging from the interaction of vector pairs, light-light (with light$\equiv \rho, \omega, \phi, K^*$), light-heavy (with heavy$\equiv D^*$, $D_s^*$ and $J/\psi$), and heavy-heavy, was done in Ref.~\cite{Molina:2009ct} and several states were obtained that could be associated to known states. In particular, as relevant for the present work, a state $0^+(2^{++})$ at 4169~MeV was obtained, coupling mostly to $D_s^* \bar{D}_s^*$, that qualifies as a $D_s^* \bar{D}_s^*$ molecule and was associated to the $X(4160)$, not the $X(4140)$. The dynamics used in Ref.~\cite{Molina:2009ct} is based on the local hidden gauge approach~\cite{Bando:1984ej,Bando:1987br,Meissner:1987ge}, exchanging vector mesons and including contact terms. The channels included in the interaction are $D^* \bar{D}^*$,  $D_s^* \bar{D}_s^*$, $K^* \bar{K}^*$, $\rho\rho$, $\omega\omega$, $\phi\phi$,  $J/\psi J/\psi$ $\omega J/\psi$, $\phi J/\psi$, $\omega\phi$. It was shown there that the couplings of the state to the $D_s^*\bar{D}_s^*$ was dominant ($g_{D_s^*\bar{D}_s^*}=18927-5524i$~MeV), followed by the one to $\phi J/\psi$ ($g_{J/\psi \phi}=-2617-5151i$~MeV). The coupling to $D^*\bar{D}^*$ is $g_{D^*\bar{D}^*}=1225-490i$, sizeable enough, that guarantees that this resonance can be seen in the $D^*\bar{D}^*$ channel. Actually, this is the channel where the $X(4160)$ was observed in the $e^+e^- \to J/\psi X$, $X\to D^*\bar{D}^*$ reaction~\cite{Abe:2007sya}. The width of the $X(4160)$ is given by $\Gamma=139^{+119}_{-61}\pm 21$~MeV, much wider than that of the $X(4140)$. The work of Ref.~\cite{Molina:2009ct} gives $\Gamma=132\pm 25$~MeV. It should be noticed that with the coupling $g_{J/\psi \phi}$ obtained, one obtains a partial decay width $\Gamma_{J/\psi \phi}\approx 22$~MeV. So, much of the width comes from other channels, in particular the light-light ones that have much phase space for the decay.

It is interesting to note in retrospective that the theoretical works discussed above that associated the $D_s^*\bar{D}_s^*$ structure to the $X(4140)$ could equally have associated it to the $X(4160)$. One can guess that the fact that light-light  channels were not considered rendered the width of the state small and the association to the $X(4140)$ state was more natural. Yet, today, with the quantum numbers of the $X(4140)$ established to be $0^+(1^{++})$~\cite{Aaij:2016nsc,Aaij:2016iza,PDG2016}, the association of the $0^{++}$, $2^{++}$ states to the $X(4140)$ can no longer be supported, but the idea of the $D_s^*\bar{D}_s^*$ molecule associated to the $X(4160)$ has much weight. 

The dominant terms of the interaction in Ref.~\cite{Molina:2009ct} correspond to the exchange of light vectors, and they can be obtained from the picture where the heavy quarks of the vector components are spectators, and only light quarks are operative in the interaction~\cite{Sakai:2017avl}. This allows to obtain this interaction from the SU(3) light sector and at the same time, it guarantees that heavy quark symmetry is fulfilled since the heavy quarks are spectators in the interaction. Transitions from heavy-heavy to light-light require the exchange of heavy vectors leading to sub-dominant terms in the inverse of the heavy quark mass counting, which are calculated using SU(4) symmetry and one accepts as being model dependent. From this perspective, we will allow the total width of the $X(4160)$ coming from the light-light channels to be somewhat different than the one obtained in Ref.~\cite{Molina:2009ct} when we fit the data of Refs.~\cite{Aaij:2016nsc,Aaij:2016iza}.

Next we proceed to apply our approach to the data of Refs.~\cite{Aaij:2016nsc,Aaij:2016iza} which we consider from threshold up to about $4250$~MeV, above the $D_s^*\bar{D}_s^*$  threshold. The data show a narrow peak around $4140$~MeV, followed by one broader structure around $4160\sim 4170$~MeV, and a remarkable cusp structure around the $D_s^*\bar{D}_s^*$  threshold. The presence of a cusp at the $D_s^*\bar{D}_s^*$ threshold in the $J/\psi \phi$ mass distribution clearly indicates a link of the resonance responsible for the $J/\psi \phi$ spectrum with the $D_s^*\bar{D}_s^*$ channel. This link can be provided assuming that the $X(4160)$ is mostly responsible for this spectrum.

\begin{figure}
\includegraphics[width=0.35\textwidth]{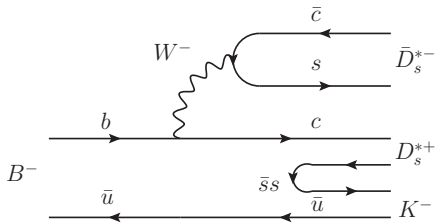}
\caption{Microscopic quark level production of $K^- D_s^*\bar{D}_s^*$ in $B^-$ decay.}  \label{fig:weakdecay}
\end{figure}

\begin{figure}
\includegraphics[width=0.4\textwidth]{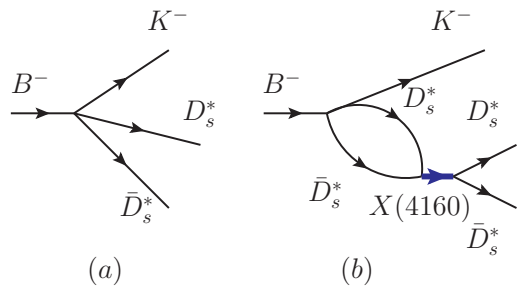}
\caption{Mechanism for $B^- \to K^- D_s^*\bar{D}_s^*$ in the presence of the $X(4160)$ resonance.}  \label{fig:DsDs}
\end{figure}

\begin{figure}
\includegraphics[width=0.35\textwidth]{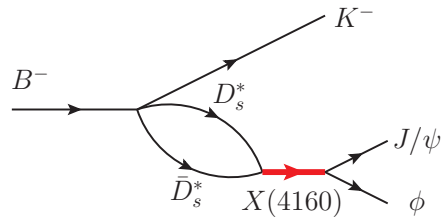}
\caption{Mechanism for $B^- \to K^- J/\psi \phi$ driven by the $X(4160)$ resonance.}
\label{fig:jpsiphi_FSI}
\end{figure}

The next step requires to discuss how a $D_s^*\bar{D}_s^*$  resonance is produced in the weak decay $B^- \to K^- D_s^*\bar{D}_s^*$ (we take the complex conjugate reaction to deal with $b$ quark rather than $\bar{b}$ quark). The dominant process at the quark level proceeds as shown in Fig.~\ref{fig:weakdecay}, involving external emission~\cite{Chau:1982da}. This allows us immediately to obtain the $B^- \to K^- D_s^*\bar{D}_s^*$ amplitude in the region around the $X(4160)$ resonance as depicted in Fig.~\ref{fig:DsDs}. Obviously in the neighborhood of the resonance the tree level term of Fig.~\ref{fig:DsDs}(a) is small compared to the resonant term, but we keep it in the calculations. For the production of $J/\psi \phi$ with this mechanism, the tree level of Fig.~\ref{fig:DsDs}(a) does not contribute and then we are led to the diagram of Fig.~\ref{fig:jpsiphi_FSI}.

\begin{figure}
\includegraphics[width=0.35\textwidth]{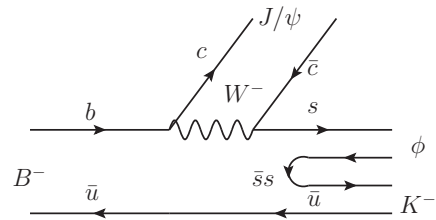}
\caption{Tree level diagram  to produce the $B^- \to K^- J/\psi \phi$ decay involving internal conversion.}
\label{fig:jpsiphi_tree}
\end{figure}

The $B^- \to K^- J/\psi \phi$ reaction can also proceed through the mechanism of Fig.~\ref{fig:jpsiphi_tree} involving internal conversion. Yet, the internal conversion is penalized by color factors with respect to the external emission of Fig.~\ref{fig:weakdecay}, and hence this term, or rescattering of this term like that in Fig.~\ref{fig:jpsiphi_FSI}, but with $J/\psi \phi$ intermediate state instead of
 $D_s^*\bar{D}_s^*$, which would involve the extra factor $g_{J/\psi\phi}/g_{D_s^*\bar{D}_s^*}$ versus the amplitude of Fig.~\ref{fig:jpsiphi_FSI}, can be safely neglected. 

We can then write the amplitude for the $B^-\to K^-J/\psi\phi$ and $B^-\to K^-D_s^*\bar{D}_s^*$. The resonance $X(4160)$ obtained in Ref.~\cite{Molina:2009ct} is a $J^{PC}=2^{++}$ state with $L=0$ in $D_s^*\bar{D}_s^*$. To match angular momentum in the reaction, we need a $D$-wave in the $K^-$ and the amplitude is thus of the type,
\begin{eqnarray}
t^{\rm tree}_{B^-\to K^- D_s^*\bar{D}_s^*} &=& A \left( \vec{\epsilon}\cdot \vec{k}\, \vec{\epsilon}^{\,\prime}\cdot \vec{k}-\frac{1}{3}\vec{k}^2\vec{\epsilon}\cdot \vec{\epsilon}^{\,\prime}\right), \label{eq:pola}
\end{eqnarray}
where $\vec{\epsilon}$, $\vec{\epsilon}^{\,\prime}$ are the polarization vectors of $D_s^*$ and $\vec{D}_s^*$, and we evaluate it in the frame of reference where the $D_s^*\bar{D}_s^*$ system is at rest.  $A$ in Eq.~(\ref{eq:pola}) is an unknown factor that will be fitted to the data. The sum over polarizations of $|t|^2$ is,
\begin{equation}
\sum_{\rm pol} |t^{\rm tree}_{B^-\to K^- D_s^*\bar{D}_s^*}|^2=\frac{2}{3}\vert \vec{k}\vert ^4 ,
\end{equation}  
and then for the tree level term of Fig.~\ref{fig:DsDs}(a) we obtain,
\begin{eqnarray}
\frac{d\Gamma}{dM_{\rm inv}(D_s^*\bar{D}_s^*)} &=& \frac{1}{(2\pi)^3}\frac{1}{4M_{B^-}^2}\frac{2}{3}|\vec{k}|^4|\vec{k}'|\, \tilde{p}_{D_s^*}|A|^2, \label{eq:dwidth}
\end{eqnarray}
where $\vec{k}$, as mentioned before, is the $K^-$ momentum in the $D_s^*\bar{D}_s^*$ rest frame, $\vec{k}'$ the $K^-$ momentum in the $B^-$ rest frame, and $\tilde{p}_{D_s^*}$ the $D_s^*$ momentum in the $D_s^*\bar{D}_s^*$ rest frame.

If we want to get the mass distribution for the mechanisms depicted in Fig.~\ref{fig:DsDs} for $D_s^*\bar{D}_s^*$ production including the $X(4160)$ resonance, we make the following replacement,
\begin{eqnarray}
A\rightarrow && A\left[ 1+G_{D_s^*\bar{D}_s^*}(M_{\rm inv}(D_s^*\bar{D}_s^*)) \right. \nonumber \\
&&\times \left. t_{D_s^*\bar{D}_s^*\to D_s^*\bar{D}_s^*}(M_{\rm inv}(D_s^*\bar{D}_s^*))\right].
\label{eq:tampDsDs}
\end{eqnarray}
To obtain the mass distribution for $J/\psi\phi$ through the mechanism of Fig.~\ref{fig:jpsiphi_FSI}, we make the following replacement in Eq.~(\ref{eq:dwidth}),
\begin{eqnarray}
A\rightarrow && A\times G_{D_s^*\bar{D}_s^*}(M_{\rm inv}(J/\psi\phi))\nonumber \\
&& \times  t_{D_s^*\bar{D}_s^*\to J/\psi\phi}(M_{\rm inv}(J/\psi\phi)), \label{eq:tampjpsiphi}
\end{eqnarray}
and and $k$, $k'$ are the same $K^-$ momenta as before suited to the situation of having $J/\psi\phi$ in the final state rather than $D_s^*\bar{D}_s^*$. Similarly, $\tilde{p}_{D_s^*}$ in Eq.~(\ref{eq:dwidth}) is replaced by $\tilde{p}_{\phi}$, which is the $\phi$ momentum in the $J/\psi\phi$ rest frame. 

The $G$ function appearing in Eqs.~(\ref{eq:tampDsDs}) and (\ref{eq:tampjpsiphi}) is the loop function for two intermediate $D_s^*\bar{D}_s^*$ . To avoid potential dangers using the dimensional regularization as pointed out in Ref.~\cite{Wu:2010rv}, we use the cut off method with $q_{\rm max}$ fixed such as to give the same value as $G$ with dimensional regularization used in Ref.~\cite{Molina:2009ct} at the pole position.

The amplitudes appearing in Eqs.~(\ref{eq:tampDsDs}) and (\ref{eq:tampjpsiphi}) are given in terms of the  $g_{D_s^*\bar{D}_s^*}$ and $g_{J/\psi\phi}$ obtained in Ref.~\cite{Molina:2009ct} by,
\begin{eqnarray}
t_{D_s^*\bar{D}_s^*\to D_s^*\bar{D}_s^*} & =& \frac{g^2_{D_s^*\bar{D}_s^*}}{M^2_{\rm inv}(D_s^*\bar{D}_s^*)-M^2_X+iM_X\Gamma_X} ,\label{eq:tDsDs} \\
t_{D_s^*\bar{D}_s^*\to J/\psi\phi} & =& \frac{g_{D_s^*\bar{D}_s^*}g_{J/\psi\phi}}{M^2_{\rm inv}(J/\psi\phi)-M^2_X+iM_X\Gamma_X}, \label{eq:tJpsiphi}
\end{eqnarray}
where,
\begin{equation}
\Gamma_X=\Gamma_0 + \Gamma_{J/\psi\phi} +\Gamma_{D_s^*\bar{D}_s^*},
\end{equation}
with $\Gamma_0$ accounting for the channels of Ref.~\cite{Molina:2009ct} not explicitly considered here (we shall fit that to the data as discussed above), and,
\begin{eqnarray}
\Gamma_{J/\psi\phi} &=&\frac{|g_{J/\psi\phi}|^2}{8\pi M^2_X}\tilde{p}_{\phi},\label{eq:dwJpsiphi} \\
\Gamma_{D_s^*\bar{D}_s^*} &=&\frac{|g_{D_s^*\bar{D}_s^*}|^2}{8\pi M^2_X}\tilde{p}_{D_s^*}\Theta(M_{\rm inv}(D_s^*\bar{D}_s^*)-2M_{D_s^*}). \label{eq:dwDsDs}
\end{eqnarray}

Eqs.~(\ref{eq:tDsDs}) and (\ref{eq:tJpsiphi}) together with Eqs.~(\ref{eq:dwJpsiphi}) and (\ref{eq:dwDsDs}) incorporate the Flatt\'e effect for the opening of the important $D_s^*\bar{D}_s^*$ channel above the $D_s^*\bar{D}_s^*$  threshold.

To account for the production of $J/\psi\phi$ via the $1^{++}$ $X(4140)$ resonance, we take the suitable operator with the kaon in $P$-wave $(\vec{\epsilon}_{J/\psi} \times \vec{\epsilon}_\phi) \cdot \vec{k}$, and the mass distribution coming from this source is given by Eq.~(\ref{eq:dwidth}) with the following substituion,
\begin{gather}
M_{\rm inv}(D_s^*\bar{D}_s^*)  \rightarrow M_{\rm inv}(J/\psi\phi), \nonumber \\
\frac{2}{3}|\vec{k}|^4 \rightarrow 2|\vec{k}|^2, ~~ \tilde{p}_{D_s^*} \rightarrow \tilde{p}_\phi, \nonumber \\
A\rightarrow \frac{B \, M^4_{X(4140)}}{M^2_{\rm inv}(J/\psi\phi)-M^2_{X(4140)}+iM_{X(4140)}\Gamma_{X(4140)}},
\end{gather}
with $B$ a parameter to  be fitted to the data and for $\Gamma_{X(4140)}$ we take the average of the PDG. Here we take $M_{X(4140)}=4135$~MeV, since this is the peak of the $X(4140)$ structure used in Refs.~\cite{Aaij:2016nsc,Aaij:2016iza}. In Refs.~\cite{Aaij:2016nsc,Aaij:2016iza} the authors use both a Breit-Wigner and a structure incorporating the $D_s\bar{D}_s^*$ threshold, as suggested in Ref.~\cite{Ashtekar:2016ecx}.

\begin{figure}
\includegraphics[width=0.4\textwidth]{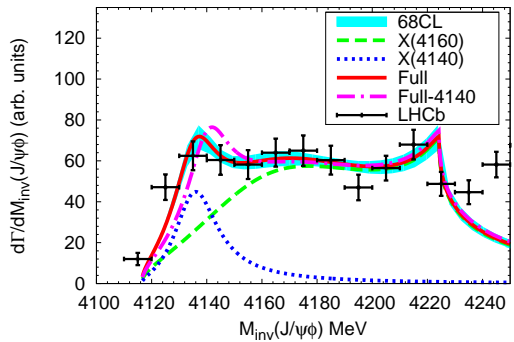}
\caption{The $J/\psi\phi$ mass distribution of the  $B^- \to K^- J/\psi \phi$ decay. The band comes considering the errors in $A$, $B$ and $\Gamma_0$ from the fit, and represents the 68\% confidence-level.}
\label{fig:result_Jpsiphi}
\end{figure}

The freedom to fit the data are the parameters $A$, $B$, and $\Gamma_0$. A suitable fit to the data is obtained as shown in Fig.~\ref{fig:result_Jpsiphi} with $\Gamma_0=67.0\pm9.4$~MeV (at 68\% confidence-level), which together with $\Gamma_{J/\psi\phi}\simeq 22.0$~MeV would provide $\Gamma_{X(4160)}\simeq 89.0\pm9.4$~MeV which is compatible with the experimental width from the PDG of $\Gamma=139^{+111}_{-61}\pm 21$~MeV. As we can see in the figure, we obtain a contribution from the $X(4140)$ (blue dotted curve) that is dominant at low invariant masses, and is responsible for the peak observed in the experiment around 4135~MeV. The $X(4160)$  (green dashed curve) is responsible for most of the strength and produces a broader peak around 4170~MeV (we take the mass $M_{X(4160)}=4169$~MeV as obtained in Ref.~\cite{Molina:2009ct}). And finally a cusp appears at the $D_s^*\bar{D}_s^*$ threshold as it shows up in the experiment. This cusp comes from the factor $G_{D_s^*\bar{D}_s^*}(M_{\rm inv})$ and reflects the analytical structure of this function with a discontinuity of the derivative at threshold. One must stress that this factor appears here as a consequence of the $D_s^*\bar{D}_s^*$ molecular structure of the $X(4160)$. In an analysis like the one of Refs.~\cite{Aaij:2016nsc,Aaij:2016iza}, where a sum of amplitudes for resonance excitation and some background are fitted to the data, this factor is not considered, and as a consequence the cusp around $D_s^*\bar{D}_s^*$ in the data is missed in the fit.

\begin{figure}
\includegraphics[width=0.4\textwidth]{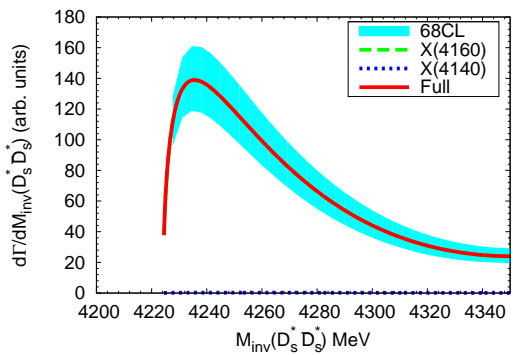}
\caption{The $D_s^*\bar{D}_s^*$ mass distribution of the $B^- \to K^- D_s^*\bar{D}_s^*$ decay. The band comes considering the errors in $A$, $B$ and $\Gamma_0$ from the fit, and represents the 68\% confidence-level.}
\label{fig:result_DsDs}
\end{figure}

The Flatt\'e effect is visible in Fig.~\ref{fig:result_Jpsiphi} as a sharp fall down of the invariant mass distribution above the $D_s^*\bar{D}_s^*$ threshold. From there on, our fit starts diverting from experiment, but so it should, since definitely, other contributions from resonances and backgrounds  as discussed in Refs.~\cite{Aaij:2016nsc,Aaij:2016iza}, should be considered. Our point is that the lower part of the spectrum can be obtained from the contribution of the $X(4140)$ $(1^{++})$ and $X(4160)$ $(2^{++})$ resonances, the $X(4140)$ is narrow, like determined in most experiments, and the peak of the distribution at the $D_s^*\bar{D}_s^*$ threshold indicates that the resonance to which the $J/\psi\phi$ is coupled in that region is strongly tied to the $D_s^*\bar{D}_s^*$ channel.

We have conducted the same fit using $M_{X(4140)}=4140$~MeV (pink dash-dotted curve in Fig.~\ref{fig:result_Jpsiphi}), and the fit  is acceptable but slightly worse in the first two points of the spectrum.

To finish the work, and as a test of the explanation given here, we present in Fig.~\ref{fig:result_DsDs} the $D_s^*\bar{D}_s^*$ mass distribution above threshold obtained with the same parameters as in Fig.~\ref{fig:result_Jpsiphi}. This should allow a quantitative comparison with the $J/\psi\phi$ mass distribution of Fig.~\ref{fig:result_Jpsiphi}, once this experiment is done, which we very much encourage.

As we can see in Fig.~\ref{fig:result_DsDs}, there is a peak close to threshold, which should not be mis-identified with a new resonance, but it is the reflection of the $X(4160)$ which in our fit has the mass at 4169~MeV. The strength at the peak is about twice the one of the X(4140) in the $J/\psi\phi$ distribution, which guarantees its observability. The $B^- \to K^- D_s^*\bar{D}_s^*$ mode is not reported in the PDG, but many modes with one $D_s^*$ already exist. The present work and the prediction, tied to the interpretation given for the $B^-\to K^-J/\psi\phi$ spectrum, should act as an incentive to measure this reaction and learn about properties of the $X(4140)$ and $X(4160)$.

\section*{Acknowledgements}

One of us, E.O, wishes to acknowledge the
support from the Chinese Academy of Science in the Program of 
CAS president's International Fellowship for visting scientists.
This work is partly supported by the National Natural
Science Foundation of China under Grant Nos. 11475227,
 11375024, 11522539, 11735003, 11505158, 11475015, and  11647601.  It is also supported by
the Youth Innovation Promotion Association CAS (No.
2016367),
the China Postdoctoral Science Foundation under Grant No. 2015M582197, the Postdoctoral Research Sponsorship in Henan Province under Grant No. 2015023, and Academic Improvement Project of Zhengzhou University.
This work is also partly supported by the Spanish Ministerio de Economia 
y Competitividad
and European FEDER funds under the contract number FIS2014-57026-REDT, 
FIS2014-51948-C2-1-P, and FIS2014-51948-C2-2-P, and the Generalitat 
Valenciana in the program Prometeo II-2014/068.

\end{document}